# Towards Feminist Intersectional XAI: From Explainability to Response-Ability

**Goda Klumbytė**
Faculty of Electrical Engineering and
Computer Science
University of Kassel, Germany
goda.klumbyte@uni-kassel.de

**Hannah Piehl**
Faculty of Electrical Engineering and
Computer Science
University of Kassel, Germany
hannah.piehl@stud.uni-frankfurt.de

**Claude Draude**
Faculty of Electrical Engineering and
Computer Science
University of Kassel, Germany
claude.draude@uni-kassel.de

## Abstract
This paper follows calls for critical approaches to computing and conceptualisations of intersectional, feminist, decolonial HCI and AI design and asks what a feminist intersectional perspective in HCXAI research and design might look like. Sketching out initial research directions and implications for explainable AI design, it suggests that explainability from a feminist perspective would include the fostering of response-ability – the capacity to critically evaluate and respond to AI systems – and would centre marginalised perspectives.

## Author Keywords
Feminist intersectionality; human-centred explainable AI; feminist XAI.

## CSS Concepts
• **Human-centered computing** • **Human computer interaction (HCI)** • **HCI theory, concepts and models**

## Introduction
The recent decade in HCI and AI research has witnessed an increase in critical interdisciplinary approaches: calls have been voiced for feminist intersectional HCI [4, 5, 7, 34, 44] and AI/machine learning [11, 31, 49, 51], post- and decolonial HCI [6,

**Theories of situated knowledge and standpoint** espouse the notion of intersectional understanding of social positioning and structural inequalities. Marginalised viewpoints offer specific perspectives with regards to how oppression, power and privilege operate and can constitute a locus of alternative knowledge practices. Standpoint theories also urge researchers to ask, from whose perspective and for whose benefit specific research is conducted and requires putting the needs and benefits of those marginalised to the fore.

The contextual orientation of the **sociotechnical turn** in HCXAI – encouraging to pay close attention to the organisational and sociotechnical context of AI in support of building XAI systems, allowing space for critical interpretation and reflexivity from the user's side – overlaps with feminist epistemological positionings.

29, 42] and AI [1, 37], disability-centred AI [18, 51], critical race perspectives in HCI [41, 45] and AI [20, 35]. This paper contributes to these calls by asking what a feminist intersectional approach to explainable AI (XAI) might look like.

We argue that feminist intersectional perspective in XAI resonates with human-centred XAI's (HCXAI) focus on values, interpersonal dynamics, socially situated nature of AI and building towards a reflective sociotechnical practice [12, 13, 14, 15, 16]. It contributes a further perspective of *social justice-centred normative orientation* of XAI that requires attention be paid to power differences and structural inequalities, challenging rationalist, universalist modalities of XAI and fostering response-ability as a capacity by users to critically evaluate and respond to AI systems. This conceptual perspective also has pragmatic implications for XAI design by engaging questions of user diversity and power dynamics in the XAI ecosystem.

## Feminist epistemology and feminist perspectives in (HC)XAI

Feminist epistemology is rooted in theories of *situated knowledge* and *standpoint*, which recognise (scientific) knowledge as situated in specific material, disciplinary, embodied, cultural, socio-political and historical contexts [21]. Knowledge is understood therefore as always partial. Partial, situated knowledge in turn offers a more grounded notion of truth and enables accountability since knowledge claims that can be located can also be made accountable to people and contexts that the claim affects or concerns. Thus, instead of value-neutral, universal objectivity situated knowledges offer "objectivity as positioned rationality" [21, p.590] and plurality of grounded perspectives.

Feminist standpoint theories argue that knowledge is not only produced from specific social positionings but also that it is heavily value-laden. They suggest that to move towards equity and justice, marginalised standpoints are to be prioritised in knowledge making endeavours [22, 23, 24, 25, 26, 27]. Standpoints or social positionings are not one-dimensional because multiple axes of categorical belonging (gender, race, class, sexuality, dis/ability, etc.) and resulting privileges or oppressions intersect in these positionings. This generates specific forms of oppression and inequality, whereby intersecting oppressions produce specific forms of discrimination [8], requiring an intersectional understanding of fairness [17] and understanding of users as occupying the loci of multi-dimensional identities.

Previous work has documented a *sociotechnical turn* in HCXAI and outlined an agenda that argues for the need to focus on values, interpersonal dynamics and socially situated nature of AI systems, building towards a reflective sociotechnical practice [14]. Researchers have highlighted that XAI depends on context as there are different user groups of XAI – such as novice users, data experts and AI experts [38] – with different explanation goals and evaluation criteria for what counts as a good explanation [39, 40, 50]. This parallels the concept of intersectionality premised on the idea that identities and social positionings are multiple and intersecting, and with the perspective of situated knowledges and standpoint theory highlighting the embeddedness and context-dependence of knowledge claims. Singh et al. [46] advocate foregrounding XAI perspectives in the Global South, challenging white Western rationalist explanation frameworks and ask what explainability would look like

**Paying attention to power and structural inequality** means asking questions such as:

- For whose benefit are the explanations generated and from whose perspective?
- To whom will they be accessible and useful?
- What is left unexplained?

In the context of XAI these questions should be asked first and foremost by designers, but they can be beneficial for all stakeholders.

for marginalised communities in rural India. Feminist intersectional calls to interrogate and challenge power hierarchies in knowledge production align with this position too.

A few papers argue for explicitly feminist perspectives in XAI [19, 28, 32, 33]. Hancox-Li et al. [19] interrogate epistemic values in feature importance methods, showing that feminist epistemological principles of plurality, attention to social context and privileging of subjugated perspectives can inform XAI by orienting it towards incorporating subjugated perspectives. Huang et al. [28] also argue for context-sensitivity, highlight the need for interdisciplinarity in XAI design to enrich "interpretive resources" and background assumptions for building better explanations and identifying bias, and call for more participation of diverse stakeholder groups. Singh et al. [47] mention intersectionality and carry out a more user-centred approach focusing on context, considering framework of explanation and its appropriateness.

## Towards feminist intersectional XAI

We offer a sketch of feminist intersectional XAI by outlining a set of perspectives for XAI and subsequent implications for XAI design. We present this sketch as an overall approach that can help conceptually, theoretically and normatively ground XAI research and design in orientation towards social justice.

### Feminist principles for XAI

*Normative orientation towards social justice and equity.* Feminist epistemology argues that knowledge, technology design and technological practice are not value neutral but rather products of sociotechnical realities that reflect specific values and agendas. Feminist XAI would foreground XAI design that contributes to justice and equity, making XAI inextricably linked to fairness, accountability, transparency and ethics (FATE). A feminist position suggests that simple attentiveness to inequality is not enough and instead AI systems and explanatory efforts should be directed at active erosion of inequalities and contribution towards just and equitable sociotechnical AI ecosystems.

*Attention to power and structural inequalities.* XAI research and design should also entail researching, understanding and accounting for the ways that concepts pertaining to explainability and explanation design are entangled with power dynamics in specific sociotechnical contexts. This entanglement can manifest in different ways: as reproduction of power dynamics and existing inequalities, as introduction of new social tensions and inequalities, or active erosion of these.

*Challenging universalist and traditional rationalist modalities of explanation.* Since feminist epistemology and intersectionality endorses plurality of perspectives and breaks with an idea of universal rationality, this implies that modalities of explanation focused exclusively on generating cognitive explanations of AI systems are not enough. To the contrary, as Black feminist, postcolonial, decolonial and Indigenous perspectives show, Western forms of rationality are culturally specific. Therefore, feminist XAI approaches should aim to expand what is meant by explanation and in what ways it can be delivered.

*Centring marginalised perspectives.* It is important to find ways to centre marginalised perspectives in XAI,

**Rationalist modalities of explanation** rely on the understanding of rationality as the basis of explanations. They imply that rationality is universal and therefore same rationality exists across different contexts and for different people. The idea of universal rationality has been critiqued because of its reductive understanding of personhood and compatibility with dehumanisation of certain groups by labelling them as "less rational" (historically those groups have included women, racialised people, LGBTQ people and other marginalised groups) [36].

**Response-ability** entails asking to which communities, which user groups, what kind of bodies are XAI researchers and designers accountable to, and how can XAI be designed in ways that creates more possibilities for these communities and groups to critically and constructively respond in the interaction with AI [10].

i.e., through encouraging participatory and inclusive processes in XAI design, fostering interdisciplinarity, and actively pursuing XAI that benefits marginalised stakeholders and their needs. This also requires generating XAI systems that actively empower users to critically interpret, evaluate and, importantly, respond towards AI systems. Such ***response-ability*** is what feminist situated knowledge theory positions as the key framework for evaluating accountability [2, 3, 21].

*Implications for XAI design*
- XAI designers could espouse *context-oriented inquiry* of the goals, means and ends of XAI design, with special attention to power dynamics and structural inequalities in the specific AI application domain and/or situation.
- This requires designers to pay attention to the *systemic level*, i.e., not only explaining the model itself but also understanding its context and the broader rationale of the specific AI system. The boundary work of deciding where a specific AI system and its effects begin and end also requires such systemic perspective, eased by having greater disciplinary diversity in the team.
- The notion of *proactivity* and the importance of focusing on intersectional perspectives from the start is crucial for feminist XAI design throughout (and beyond) all design stages [9, 42, 49].
- *Participatory and interactive* explanation design is important to understand and account for marginalised perspectives and foster critical response-ability, as well as diversifying XAI design teams and aiming at applying interdisciplinary perspectives (e.g., knowledges of social sciences and humanities) towards setting design goals and evaluation tools to better support operationalising the needs of marginalised positions.
- *Integrating intersectionality* in the understanding of different stakeholders and user groups can also be fostered by more participative design methods to address the feminist principle of centring subjugated knowledges.
- *Expanding explanatory modalities* means experimenting with the ways explanations are delivered beyond more traditional rational-cognitive explanations and is another important implication.

## Discussion
To end, we offer some questions for discussion and further research. While feminist intersectional perspective for XAI can provide broader orientation, to what extent can it be formalised into concrete application methods, and even then, to what extent can these methods be generalised for different situations? We believe feminist intersectional XAI would necessitate resisting scalability and instead focusing on case-specificity [30, 48]. To what extent can feminist intersectional XAI be applied in more commercial, profit oriented XAI systems? Since feminist intersectional perspective is inherently political and normatively oriented, is a broader uptake possible and what might be required for it? Can a feminist intersectional perspective in XAI also be productively engaged by decolonial, postcolonial and other non-Eurocentric perspectives, or does it remain constrained by its inevitable rootedness in specific (Global Northern) situatedness? We suggest that feminist intersectional XAI is a domain open to and in need of further investigation.

## Acknowledgements

This research is supported by Volkswagen Foundation grant "Artificial Intelligence and the Society of the Future" as part of the collaborative project "AI Forensics: Accountability through Interpretability in Visual AI Systems".